\documentclass[aps,prc,10pt,twocolumn]{revtex4-2}
\usepackage{amssymb,amsmath}
\usepackage{graphicx}
\usepackage{color}

\renewcommand{\figurename}{Fig.}

\begin{document} 

\title{Relaxation of a single-particle excitation in a Fermi system within the diffusion approximation of kinetic theory}

\author{Sergiy V. Lukyanov}

\email{lukyanov@kinr.kiev.ua}

\affiliation{\it{Institute for Nuclear Research, 03680 Kyiv, Ukraine}}

\date{\today}

\begin{abstract}
The time evolution of the Wigner distribution function for a single-particle excitation in a Fermi system was studied within the framework of the diffusion approximation of kinetic theory by numerically solving a nonlinear diffusion equation with constant kinetic coefficients. A method was proposed to separate the dissipative processes into contributions from the relaxation of the single-particle excitation and from the relaxation of the nuclear core, with a distinct relaxation time introduced for each process. The influence of the diffusion and drift coefficients on the characteristic relaxation timescale was analyzed. It was found that the resulting relaxation times exhibit a discrepancy relative to the kinetic coefficient estimates known from previous studies.
\end{abstract}

\maketitle

\section{Introduction}

The study of relaxation processes in Fermi systems is one of the key directions in modern theoretical physics, as these processes determine the time evolution of 
nonstationary states, the transport properties of matter, and the mechanisms by which thermodynamic equilibrium is established. Such processes manifest themselves 
across a wide range of phenomena—from the thermalization of excitations in heavy-ion collisions to the kinetics of collective oscillations in nuclei and 
the description of dense nuclear matter in astrophysical objects.

To describe the evolution of nonstationary states, the Landau–Vlasov kinetic equation is commonly employed \cite{KoSh.b.2020}. This approach significantly simplifies 
the description of the system; however, the collision integral on the right-hand side of the equation remains a complicated analytical object that requires the use of 
specific approximations. One of these is the so-called diffusion approximation \cite{LiPi.bp2.1981}, within which collisions are represented as effective diffusion and 
drift processes in momentum space. This approach makes it possible to study the relaxation properties of the system in a generalized way without resorting to a detailed 
consideration of the microscopic collision mechanism.

The evolution of a Fermi system is determined by the kinetic coefficients of diffusion and drift that appear in the diffusion equation. These coefficients govern 
the intensity of dissipative processes of both a collective and a single-particle nature. Knowledge of their magnitudes provides information not only about the relaxation time 
but also about the equilibrium temperature of the system. For example, in Ref. \cite{Wo.PRL.1982} a schematic model was proposed to describe the equilibrium state of a finite 
Fermi system. It was shown that the damping time of such an excitation is determined by the kinetic coefficients of the system and is given by $\tau_\text{equ} = 4D/v^2$, 
where $D$ is the diffusion coefficient and $v$ is the drift coefficient. The equilibrium temperature of the system is likewise expressed through a simple relation 
between these coefficients: $T_\text{eq} = -D/v$.

In Refs. \cite{KoLu.UPJ.2014,KoLu.IJMP.2015}, explicit expressions for the diffusion and drift coefficients were obtained within the low-temperature approximation. 
However, it was found that the assumption of small transferred momenta in particle scattering near the Fermi surface \cite{AbKh.RPP.1959} is insufficient to yield 
physically consistent results. Previously, the approximation of an isotropic nucleon scattering probability had been considered satisfactory 
\cite{AbKh.RPP.1959,KoLuPlSh.PRC.1998}. Nevertheless, when the kinetic coefficients are calculated under this assumption, divergent values arise in the integral expressions. 
The problem was resolved by imposing a short-range constraint on the interparticle potential. In particular, for a Gaussian potential, convergent expressions for 
the kinetic coefficients were obtained, and their numerical values were computed solely from phenomenological parameters of the interparticle interaction—namely, 
the potential depth and its effective range \cite{KoLu.IJMP.2015}.

As is well known, within the framework of the Fermi-liquid drop model, the diffusion coefficient is related to the two-particle relaxation time $\tau_\text{coll}$ 
of a collective excitation, such as the giant multipole resonances in atomic nuclei \cite{We.PRB.1974,Be.ZPA.1978,Ko.NPA.1982,ShKo.RPP.2004}. Calculations have shown 
that the typical two-particle relaxation time in a nucleus is $\tau_\text{coll} \approx 10^{-23}\text{–}10^{-22}$ s. In Ref. \cite{Lu.APPB.2025}, a phenomenological method 
was proposed for evaluating the effective relaxation time of an arbitrary initial excitation as it approaches its equilibrium state. The relaxation time was defined 
as the area under the equivalent exponential time dependence of the root-mean-square deviation of the distribution function, normalized to its initial value. According 
to this approach, for a steplike initial distribution, the relaxation time was found to be $\tau_\text{eq} \approx 10^{-24}$ s. The analysis showed that the obtained relaxation 
time $\tau_\text{eq}$ is approximately $1$ order of magnitude smaller than the expected two-particle relaxation time.

At the same time, it should be noted that the effective relaxation time mentioned above characterizes the relaxation processes in the system as a whole; i.e., 
it includes both the relaxation of the single-particle excitation and that of the background.
In the present work, the two contributions are separated, and the relaxation features of each component are studied individually.
In addition, I analyze how the values of the kinetic coefficients affect the timescales of the distribution function’s evolution and the effective relaxation time.
This makes it possible to assess how variations in the model parameters can alter the characteristic relaxation times in Fermi systems.

\section{Diffusion approximation to the kinetic equation}

Consider the Landau-Vlasov kinetic equation with the collision integral on the right-hand side
\begin{equation}\label{kineq}
\frac{\partial f}{\partial t}+\hat{L}f = \mathrm{St}\{f\},
\end{equation}
where $f\equiv f(\vec{r},\vec{p},t)$ is the Wigner distribution function in phase space, $\mathrm{St}\{f\}$ is the collision integral, and the operator $\hat{L}$ 
is defined as
\begin{equation}\label{operator_l}
\hat{L}=\frac{\vec{p}}{m}\cdot\vec{\nabla}_{r}-\left[\vec{\nabla}_{r} U\right]\cdot\vec{\nabla}_{p}
\end{equation}
where $m$ is the mass of a particle. In the general case, the single-particle potential $U$ includes both self-consistent and external fields.

For the collision integral, the expression obtained earlier within the diffusion approximation is used \cite{KoLu.UPJ.2014,KoLu.IJMP.2015}
\begin{eqnarray}
\mathrm{St}\{f\}= &-&\vec{\nabla}_{p}\cdot\left[ K_p f (1-f) \frac{\vec{p}}{m} + f^2 \vec{\nabla}_{p} D_p  \right] 
\nonumber \\
&+& \vec{\nabla}_{p}^2 \left[f D_p\right].
\label{st}
\end{eqnarray}
Here, $D_p$ is the second moment with respect to the momentum transfer $\vec{s}$ in the nucleon scattering process on the Fermi surface, 
which defines the diffusion coefficient in momentum space
\begin{equation}\label{dp}
D_p=\frac{1}{6}\int\frac{\mathsf{g}d\vec{s}}{(2\pi \hbar )^{3}}\ \vec{s}^{\ 2}\ W,
\end{equation}
and the quantity $K_p$ defines the drift coefficient
\begin{equation}\label{kp}
K_p=\frac{m}{p}\left[\hat{p}\cdot\vec{\nabla}_p D_p-\int\frac{\mathsf{g}d\vec{s}}{(2\pi \hbar )^{3}}\ \hat{p} \vec{s}\ W\right],
\end{equation}
where
\begin{eqnarray}
W \approx \frac{2\mathsf{g}}{m^{2}} \frac{d\sigma }{d\Omega}(s^{2}) \int d\vec{p}_{2}\ d\vec{p}_{2}^{\ \prime}\ (1-f_2)
f_2'\ \delta \left( \vec{p}_{2}-\vec{p}_{2}^{\ \prime}-\vec{s}\right)
\nonumber \\
\times \delta\left(\epsilon_{2}-\epsilon_{2}^{\prime} - \vec{p}\cdot\vec{s}/m\right). \ \ \ \ \ \ \ \ \ \ \ \ \ \ \ \  
\label{wps}
\end{eqnarray}
Here, $\mathsf{g}=4$ is the degeneracy factor of nucleons. In Eq. (\ref{wps}), $d\sigma(s^{2})/d\Omega$ is the differential cross section for nucleon scattering, 
and $\epsilon_2=p_2^2/2m$ and $\epsilon_2^{\prime}=p_2^{\prime 2}/2m$ are the single-particle energies before and after scattering.

The study focuses on a Fermi system that models a spherical atomic nucleus in its ground state. To describe it, one uses the approximation of infinite nuclear matter.
In this case, the distribution function does not depend on spatial coordinates, $f=f(\vec{p},t)$, and the action of the operator $\hat{L}$ on the distribution function 
is zero, $\hat{L}f=0$. The distribution function in momentum space is assumed to be spherically symmetric, $f(\vec{p},t)=f(p,t)$.
For Eqs. (\ref{dp}) and (\ref{kp}), the constant kinetic coefficients \cite{Wo.PRL.1982,KoLu.UPJ.2014,KoLu.IJMP.2015,BaWo.AP.2019,Lu.NPAE.2023} are used: 
$D_p = D_{p,0}$ and $K_p = K_{p,0}$.

Taking into account the above approximations for the collision integral in Eq.~(\ref{st}), the kinetic equation in Eq.~(\ref{kineq}) transforms into the following nonlinear 
diffusion equation:
\begin{eqnarray}
\frac{\partial f}{\partial t} = &-& \frac{K_{p,0}}{m} \left[ p(1 - 2f) \frac{\partial f}{\partial p} + 3 f(1-f) \right]
\nonumber \\
&+& D_{p,0} \left[ \frac{\partial^2 f}{\partial p^2}+ \frac{2}{p} \frac{\partial f}{\partial p} \right],
\label{eqdif0}
\end{eqnarray}
which is solved numerically in Sec.~\ref{sec:numcalc}.

In this study, the values of $D_{p,0}$ and $K_{p,0}$ are treated as model parameters and may be chosen arbitrarily. At the same time, their selection is guided by 
the results of previous studies, in particular Ref.~\cite{Lu.NPAE.2023}, in which these coefficients were calculated at the Fermi surface at zero temperature:
$D_{p}^{(0)}(p_\text{F}) \approx 3.38 \times 10^{-22}$ MeV$^{2}$ fm$^{-2}$ s and
$K_{p}^{(0)}(p_\text{F}) \approx -2.76 \times 10^{-23}$ MeV fm$^{-2}$ s.

In the constant-coefficient approximation, $D_{p,0} = D_{p}^{(0)}(p_\text{F})$ and $K_{p,0} = K_{p}^{(0)}(p_\text{F})$, and their ratio, as is known 
\cite{KoLu.IJMP.2015,Lu.IJMP.2021,Wo.PRL.1982,BaWo.AP.2019}, determine the equilibrium temperature of the system:
\begin{equation}
T_\text{eq} = -\dfrac{D_{p,0}}{K_{p,0}}.
\label{tequil}
\end{equation}
Using the numerical values of the kinetic coefficients given above, one obtains $T_\text{eq} \approx 12$ MeV. This value is approximately three times higher
than the equilibrium temperature reported in Refs.~\cite{Wo.PRL.1982,BaWo.AP.2019}. Considering that the critical temperature of nuclear matter is $T_\text{C} = 18$ MeV 
\cite{RaPeLa.NPA.1983}, it is evident that the coefficients $D_{p}^{(0)}(p_\text{F})$ and $K_{p}^{(0)}(p_\text{F})$ require refinement.

Following Refs.~\cite{Wo.PRL.1982,BaWo.AP.2019}, in the subsequent calculations one uses $T_\text{eq} = 4$ MeV as the equilibrium temperature.
Then, keeping the diffusion coefficient $D_{p}^{(0)}(p_\text{F})$ unchanged, a drift coefficient renormalized by a factor of $3$ is used
in order to satisfy Eq.~(\ref{tequil}): $K_{p,0} \approx 3 K_{p}^{(0)}(p_\text{F})$.

\section{Relaxation time of single-particle excitation}

Consider the deviation of the distribution function from its equilibrium value
\begin{equation}
\delta f(p,t)=f(p,t)-f_\text{eq}(p).
\label{df}
\end{equation}
It is clear that at time $t=0$, the distribution function corresponds to the initial state
\begin{equation}
f(p,t=0)=f_\text{in}(p).
\end{equation}

In the case of a single-particle excitation, the initial distribution can be represented as the sum of a steplike distribution, $f_{\text{in},0}(p)$, 
and an excitation, $f_{\text{in},1}(p)$, which describes the single-particle peak (see Sec.~\ref{sec:numcalc})
\begin{equation}
f_\text{in}(p)=f_{\text{in},0}(p)+f_{\text{in},1}(p).
\label{fin}
\end{equation}

Let us introduce the notation for the initial deviation from the equilibrium distribution
\begin{equation}
\delta f(p,t=0)\equiv\delta f_\text{in}(p)=f_\text{in}(p)-f_\text{eq}(p).
\label{dfin}
\end{equation}
Similarly, for the steplike distribution function, one has
\begin{equation}
\delta f_{\text{in},0}(p)=f_{\text{in},0}(p)-f_{\text{eq},0}(p),
\label{dfin0}
\end{equation}
where $f_{\text{eq},0}(p)$ is the equilibrium value of the distribution function to which the steplike initial distribution $f_{\text{in},0}(p)$ evolves.
Note that $f_{\text{eq},0}(p)$ differs from the overall equilibrium distribution $f_\text{eq}(p)$ due to a different value of the Fermi energy.
Furthermore, it is clear that the evolution of the deviation from the equilibrium of the initial step-like distribution is described by the difference
\begin{equation*}
\delta f_{0}(p,t)=f_{0}(p,t)-f_{\text{eq},0}(p).
\end{equation*}

Substituting Eqs.~(\ref{fin}) and (\ref{dfin0}) into Eq.~(\ref{dfin}) and performing several transformations, one obtains
\begin{eqnarray}
\delta f_\text{in}(p)&=&f_\text{in}(p)-f_\text{eq}(p) 
\nonumber \\
&=& f_{\text{in},0}(p)+f_{\text{in},1}(p)-f_\text{eq}(p) 
\nonumber \\
&=& \delta f_{\text{in},0}(p)+f_{\text{eq},0}(p)+f_{\text{in},1}(p)-f_\text{eq}(p) 
\nonumber \\
&=& \delta f_{\text{in},0}(p)+f_{\text{in},1}(p)-\left[f_\text{eq}(p)-f_{\text{eq},0}(p)\right]. \ \ \ \ \ 
\label{dfin1}
\end{eqnarray}

The difference between the equilibrium functions, $f_\text{eq}(p)-f_{\text{eq},0}(p)$, in Eq.~(\ref{dfin1}) defines the contribution from the single-particle excitation, 
i.e., its conditional equilibrium part. Let us denote it as
\begin{equation}
f_{\text{eq},1}(p)=f_\text{eq}(p)-f_{\text{eq},0}(p).
\label{feq1p}
\end{equation}
This distribution has meaning only against the background of other equilibrium functions, and in their absence or if they are identical, it equals $0$.

Taking into account Eq.~(\ref{feq1p}), Eq.~(\ref{dfin1}) can be rewritten as
\begin{equation}
\delta f_\text{in}(p)=\delta f_{\text{in},0}(p)+f_{\text{in},1}(p)-f_{\text{eq},1}(p).
\label{dfin2}
\end{equation}

Similarly to Eqs.~(\ref{df}) and (\ref{dfin0}), one can write the expression for the deviation of the single-particle excitation from its equilibrium part
\begin{equation}
\delta f_{\text{in},1}(p)=f_{\text{in},1}(p)-f_{\text{eq},1}(p).
\label{dfin1p}
\end{equation}

Then, taking into account Eq.~(\ref{dfin1p}), Eq.~(\ref{dfin2}) takes the form
\begin{equation}
\delta f_\text{in}(p)=\delta f_{\text{in},0}(p)+\delta f_{\text{in},1}(p).
\label{dfin3}
\end{equation}

One assumes that, analogously, the time evolution of these deviations behaves in the same way
\begin{equation}
\delta f(p,t)=\delta f_0(p,t)+\delta f_1(p,t).
\label{df2}
\end{equation}

For the sake of unifying the subsequent notation, one introduces the index \text{tot} (for “total”): $\delta f(p,t) \equiv \delta f_\text{tot}(p,t)$ and
$\delta f_\text{in}(p) \equiv \delta f_{\text{in,tot}}(p)$.

Similarly to the approach used in Ref.~\cite{Lu.APPB.2025}, one considers the root-mean-square deviation, 
but this time for each deviation of the distribution function
\begin{equation}
\Delta_{n}(t) = \sqrt{\int d\vec{p}\ \left[ \delta f_{n}(p,t) \right]^2},
\label{delta-n}
\end{equation}
where $n = \text{tot},0$ and $1$. The initial values are given by
\begin{equation}
\Delta_{n}(0)= \sqrt{\int d\vec{p}\ \left[ \delta f_{\text{in},n}(p) \right]^2}.
\label{deltan0}
\end{equation}

Let us assume that for each $n$ the relaxation is described by an exponential law
\begin{equation}
\delta f_{n}(p,t)=\delta f_{\text{in},n}(p) \exp\left(-t/\tau_{n}\right),
\label{dfn-exptaun}
\end{equation}
where $\tau_\text{tot}\equiv\tau$ is the relaxation time of the entire system, $\tau_0$ corresponds to the steplike distribution, 
and $\tau_1$ corresponds to the single-particle excitation.

Then, substituting Eq.~(\ref{dfn-exptaun}) into Eq.~(\ref{delta-n}) and taking into account Eq.~(\ref{deltan0}), one obtains
\begin{equation}
\Delta_n(t)=\Delta_n(0) \exp\left(-t/\tau_n\right).
\label{delta-exp}
\end{equation}

Integrating over time
\begin{equation}
\int_0^\infty\Delta_n(t) dt=\Delta_n(0) \int_0^\infty\exp\left(-t/\tau_n\right) dt.
\label{deltan-intt}
\end{equation}
After performing the integration, this yields
\begin{equation}
\tau_n = \frac{1}{\Delta_n(0)} \int_0^\infty \Delta_n(t) dt.
\label{taun-int}
\end{equation}

Substituting Eq.~(\ref{delta-exp}) into Eq.~(\ref{taun-int}) yields an identity; however, if $\Delta_n(t)$ is not purely exponential, Eq.~(\ref{taun-int}) 
defines the corresponding effective relaxation time.

\section{Numerical calculations \label{sec:numcalc}}

The diffusion equation, Eq.~(\ref{eqdif0}), must be supplemented with an initial condition. 
As this condition, one first considers a steplike distribution function
\begin{equation}
f(p,t=0)=f_{\text{in},0}(p)=\Theta\left(p_\text{F}^2-p^2\right),
\label{fin0}
\end{equation}
where $p_\text{F}$ is the Fermi momentum, which is determined from the condition of nucleon number conservation
\begin{equation}
\frac{4\pi\mathsf{g}V}{(2\pi\hbar)^{3}}\int_{0}^{\infty} dp\ p^2 f_{\text{in},0}(p)=A.
\label{mass-cons}
\end{equation}
Here, $V$ denotes the nuclear volume. The condition in Eq.~(\ref{mass-cons}) can be rewritten in the form
\begin{equation}
\frac{4\pi\mathsf{g}}{(2\pi\hbar)^{3}}\int_{0}^{p_\text{F}} dp\ p^2 =\rho,
\label{mass-cons-1}
\end{equation}
Here, $\rho = A/V$ denotes the nuclear density. It is assumed
that the matter density equals the standard nuclear matter density $\rho = 0.16$ fm$^{-3}$.
In this case, the Fermi momentum $p_\text{F}$ obtained from the condition Eq.~(\ref{mass-cons-1})
corresponds to the Fermi energy $E_\text{F} = p_\text{F}^2 / 2m \approx 37.1$ MeV.

Figure~\ref{fig:1} shows a set of profiles of the distribution function $f_0(p,t)$, calculated from the diffusion equation, Eq.~(\ref{eqdif0}), 
at different moments in time. These profiles illustrate the time evolution of the initial steplike distribution, Eq.~(\ref{fin0}).
\begin{figure}
\begin{center}
\includegraphics[width=0.95\columnwidth,clip]{Fig1.eps}
\caption{Profiles of the distribution function $f_0(p,t)$ at the time points indicated in the captions. 
The dashed curve represents the initial steplike distribution function, Eq.~(\ref{fin0}), while the solid curve shows the equilibrium Fermi distribution function, 
Eq.~(\ref{feq}), at $t = 3.0 \times 10^{-22}$ s.}
\label{fig:1}
\end{center}
\end{figure}
The momentum is presented in units of the Fermi momentum $p/p_\text{F}$, so that $p/p_\text{F} = 1$ corresponds to the Fermi surface. The figure emphasizes the region 
of momentum space with the highest diffusivity, approximately covering the interval $0.75 \leq p/p_\text{F} \leq 1.25$. The dashed curve represents the initial step-like 
distribution [Eq. (\ref{fin0})]; 
the dash-dotted, dotted, and dot–dash–dash curves correspond to the profiles calculated at times $t_1 = 7.2 \times 10^{-26}$ s, $t_2 = 4.3 \times 10^{-25}$ s, and 
$t_3 = 1.4 \times 10^{-24}$ s, respectively. As seen in the figure, the distribution function $f_0(p,t)$ becomes progressively smeared with time and gradually 
approaches the equilibrium Fermi distribution (solid curve for $t \geq 3.0 \times 10^{-22}$ s)
\begin{equation}
f_\text{eq}(p)=\left(1+\exp\left[\frac{p^2/2m - E_{\text{F,eq}}}{T_\text{eq}}\right]\right)^{-1},
\label{feq}
\end{equation}
where $E_{\text{F,eq}} = 36.7$ MeV is the Fermi energy, which is lower than $E_\text{F}$ for the steplike distribution due to the presence of the diffusive region, 
and is determined from the condition of nucleon number conservation
\begin{equation}
\frac{4\pi\mathsf{g}V}{(2\pi\hbar)^{3}} \int_{0}^{\infty} dp\ p^2 f_\text{eq}(p) = A.
\label{massconsfeq}
\end{equation}

The calculation of the relaxation time according to Eq.~(\ref{taun-int}) for $n=0$, based on the evolution of the distribution function shown in \figurename~\ref{fig:1}, 
gives the value $\tau_{0} \approx 8.9 \times 10^{-24}$ s. This value is slightly larger than that obtained in Ref.~\cite{Lu.APPB.2025}, which results from the choice of 
a different, more realistic value for the nuclear matter density.

It should also be noted that, although in the present work one employs values of the kinetic diffusion and drift coefficients that are close to those reported
in Ref.~\cite{Wo.PRL.1982} (see also Ref.~\cite{Lu.NPAE.2023}), the resulting relaxation time $\tau_{0} \approx 8.9 \times 10^{-24}$~s is approximately three times smaller 
than the corresponding value $\tau_\text{equ} = 3.05 \times 10^{-23}$~s obtained in Ref.~\cite{Wo.PRL.1982}.

Recall that in Refs.~\cite{Wo.PRL.1982,BaWo.AP.2019} exact analytical methods for solving the diffusion equation in energy space were developed. Within this approach, 
the exponential decay law of the initial excitation, characterized by the equilibrium relaxation time $\tau_\text{equ} = 4D/v^{2}$, was obtained as the asymptotic behavior 
of the exact solution in the limit $t \to \infty$.

The observed discrepancy can be explained as follows. In my approach, according to Eq.~\eqref{taun-int}, the relaxation time is defined as the area under an equivalent 
exponential time dependence of the normalized root-mean-square deviation of the distribution function (see also Ref.~\cite{Lu.APPB.2025}). In other words, the actual time evolution 
of this deviation is effectively approximated by an exponential function.

At small times, the solution of the diffusion equation exhibits a pronounced nonexponential behavior (see, e.g., \figurename~2 in Ref.~\cite{Lu.APPB.2025}), as a result of 
which the relaxation time defined in this way has the character of an effective quantity. An advantage of the proposed approach is that the effective relaxation time is sensitive 
to the configuration of the initial state. This makes it possible to investigate the dependence of the relaxation time on the parameters characterizing the system, in particular 
on the mass number and the energy of the single-particle excitation.

With increasing time, the exact solution lies above the corresponding exponential dependence but gradually acquires an almost exponential character. Clearly, if the solution 
were fitted by an exponential function in the limit $t \to \infty$, one would obtain not an effective but a more “physical” relaxation time—this is precisely the situation 
realized in Wolschin’s approach. In this case, the relaxation time entering Eq.~\eqref{delta-exp} would have to be chosen larger in order, figuratively speaking, to raise 
the exponential curve to the numerically obtained time dependence. Although such a procedure was not performed in the present work, it is evident that under these conditions 
the value of $\tau_{0}$ would approach $\tau_\text{equ}$.

In practice, due to the physical limitations of numerical precision and the accumulation of numerical errors, the normalized root-mean-square deviation of the exact solution at 
large times gradually loses numerical accuracy and ceases to decrease asymptotically, and in some cases may even begin to increase. This leads to difficulties in the numerical 
time integration. Therefore, in Eq.~\eqref{taun-int} the upper limit of the numerical integration was restricted to $t_\text{max} = 3 \times 10^{-22}$~s. Under these conditions, 
the application of numerical procedures close in spirit to Wolschin’s method is limited by the numerical stability of the employed algorithm.

One now considers the case of a single-nucleon excitation of the nucleus, described by the following initial distribution function
\begin{eqnarray}
f_\text{in}(p) &=& \Theta \left({p'}_\text{F}^2-p^2\right) \nonumber \\
&+& \left[ 1-\Theta (p^2-p_{2}^2)\right] \Theta (p^2-p_{1}^2)\Theta \left(p^2-{p'}_\text{F}^2\right), \ \ \ \ \ 
\label{fin1p}
\end{eqnarray}
where the particle is located beyond the Fermi surface with momentum $p_1 > p'_\text{F}$, and $p'_\text{F}$ is the Fermi momentum determined from 
the nucleon number conservation condition
\begin{eqnarray}
&&\frac{4\pi\mathsf{g}V}{(2\pi\hbar)^{3}} \int_{0}^{p'_\text{F}}dp\ p^{2} f_\text{in}(p)
\nonumber \\
&&\equiv \frac{4\pi\mathsf{g}V}{(2\pi\hbar)^{3}}\int_{0}^{\infty}dp\ p^{2} f_{\text{in},0}(p)= A-1.
\label{pf}
\end{eqnarray}
Note that the Fermi momentum $p'_\text{F}$ is marked with a prime to emphasize its difference from $p_\text{F}$ used in the previous case of a steplike distribution 
in infinite nuclear matter [Eq.~(\ref{fin0})]. Dividing both sides of Eq.~(\ref{pf}) by the nuclear volume, one finds
\begin{equation}
\frac{4\pi\mathsf{g}}{(2\pi\hbar)^{3}}\int_{0}^{p'_\text{F}}dp\ p^{2} f_\text{in}(p)= \rho-\frac{1}{V},
\label{pf1}
\end{equation}
where the right-hand side includes a term inversely proportional to the volume. Hence, the Fermi momentum $p'_\text{F}$ becomes dependent on the system size.

The width of the interval in momentum space corresponding to the excited particle, $\Delta p = p_2 - p_1$, is defined by the following condition
\begin{eqnarray}
&& \frac{4\pi\mathsf{g}V}{(2\pi\hbar)^{3}}\int_{p'_\text{F}}^{\infty}dp\ p^{2} f_\text{in}(p) \nonumber \\
&& \equiv \frac{4\pi\mathsf{g}V}{(2\pi\hbar)^{3}}\int_{0}^{\infty}dp\ p^{2} f_\text{in,1}(p)=1,  
\label{fin1p-dp}
\end{eqnarray}
and is inversely proportional to the mass number and to the square of the momentum (see Ref.~\cite{Lu.APPB.2025}, Appendix A)
\begin{equation}
\Delta p \approx \frac{{p'_\text{F}}^3}{3 A p_1^2}.
\label{p-width}
\end{equation}

Using the explanations and notations introduced in the previous section, the distribution function in Eq.~(\ref{fin1p}) can be written as the sum of two components. 
The first component describes a stepwise distribution
\begin{equation*}
f_{\text{in},0}(p) = \Theta \left({p'}^2_\text{F}-p^2\right),
\end{equation*}
while the second component represents the single-particle peak
\begin{equation*}
f_{\text{in},1}(p) = \left[ 1-\Theta (p^2-p_{2}^2)\right] \Theta (p^2-p_{1}^2)\Theta \left(p^2-{p'}^2_\text{F}\right).
\end{equation*}

Figure~\ref{fig:2} shows a set of profiles of the distribution function $f(p,t)$, calculated from the diffusion equation Eq.~(\ref{eqdif0}) at different times.
These profiles illustrate the overall evolution of the initial distribution, Eq.~(\ref{fin1p}). All notations are the same as those used in \figurename~\ref{fig:1}.
\begin{figure}
\begin{center}
\includegraphics[width=0.95\columnwidth,clip]{Fig2.eps}
\end{center}
\caption{Profiles of the distribution function $f(p,t)$ at different times, as indicated in the legend.
The dashed curve corresponds to the initial distribution function, Eq.~(\ref{fin1p}), while the solid curve represents the equilibrium Fermi distribution 
function, Eq.~(\ref{feq}).}
\label{fig:2}
\end{figure}
As can be seen from the figure, the distribution function $f(p,t)$, similarly to the case of a steplike distribution, gradually smooths out over time and approaches 
the equilibrium Fermi distribution, Eq.~(\ref{feq}). The corresponding equilibrium Fermi energy $E'_{\text{F,eq}}$ is higher than the Fermi energy of the initial 
steplike distribution, $E'_\text{F} = {p'_\text{F}}^2 / 2m$. This difference arises from the one-nucleon change in the atomic number. It is also evident that, over time, 
the peak corresponding to the single-particle excitation gradually smooths out and eventually disappears.

Using Eq.~(\ref{taun-int}) for the relaxation time at $n=\text{tot}$, one calculates its dependence on the single-nucleon excitation energy $E_\text{ex}$
for three representative mass numbers $A = 50$, $150$, and $250$. The results of these calculations are presented in \figurename~\ref{fig:3}.
\begin{figure}
\begin{center}
\includegraphics[width=0.95\columnwidth,clip]{Fig3.eps}
\end{center}
\caption{Relaxation time of the nucleus, $\tau$, as a function of the single-nucleon excitation energy $E_\text{ex}$. 
Calculations are shown for mass numbers $A = 50$, $150$, and $250$.}
\label{fig:3}
\end{figure}
The solid curve shows the result for the step distribution in the case of infinite nuclear matter, whose profiles are presented in \figurename~\ref{fig:1}. 
It is evident that the relaxation time $\tau_0$ in this case does not depend on $E_\text{ex}$. For a finite-size Fermi system, an inverse dependence of 
the relaxation time $\tau$ on the excitation energy emerges. As in my previous study Ref.~\cite{Lu.APPB.2025}, it is seen that with increasing excitation energy, 
the system's relaxation time decreases nonlinearly. It is also noteworthy that as the mass number $A$ increases, the relaxation time $\tau$ grows and approaches the limiting 
value for infinite nuclear matter, which is quite natural. At the same time, as in my previous work Ref.~\cite{Lu.APPB.2025}, the relaxation times obtained in this study 
are of the order of $\tau \approx 10^{-24}$~s, which is too short and does not agree with the typical two-particle relaxation time in the nucleus, 
$\tau_\text{coll} \approx 10^{-23} \div 10^{-22}$~s, Refs.~\cite{We.PRB.1974,Be.ZPA.1978,Ko.NPA.1982,ShKo.RPP.2004}.

Now, one studies the relaxation process of a single-nucleon excitation. To this end, the single-particle excitation is separated
from the overall distribution, and its individual relaxation time is calculated according to Eq.~(\ref{taun-int}) for $n = 1$.

For a clearer understanding of the particle's behavior, \figurename~\ref{fig:4} shows a set of profiles of the deviation of the single-nucleon distribution 
$\delta f_1(p,t)$, calculated using the diffusion equation, Eq.~(\ref{eqdif0}), and constructed analogously to the previous figures.
\begin{figure}
\begin{center}
\includegraphics[width=0.95\columnwidth,clip]{Fig4.eps}
\end{center}
\caption{Profiles of the deviation of the distribution function $\delta f_1(p,t)$ at the time moments indicated in the captions. 
The dashed curve represents the initial difference $f_\text{in}(p)-f_{\text{in},0}(p)$, 
while the solid curve corresponds to the equilibrium Fermi distribution, Eq.~(\ref{feq}).}
\label{fig:4}
\end{figure}
For better visibility of the details in the figure, only the lower part of the distribution, from 0 to 0.55 on the ordinate, is shown.
The initial state (dashed curve) corresponds to the difference between the distributions $f_\text{in}(p)$ and $f_{\text{in},0}(p)$.
As seen in the figure, over time the peak of the deviation spreads out, and its maximum gradually shifts to the left—toward the Fermi surface.
In the equilibrium state (solid curve), the nucleon is localized near the Fermi surface. In other words, it loses its initial energy and transitions 
to a thermodynamic equilibrium with the other nucleons of the nucleus, with the most probable momentum corresponding to the Fermi momentum.
This process occurs over a characteristic time that can be described by the relaxation time $\tau_1$, defined according to Eq.~(\ref{taun-int}) for $n = 1$.

The results of this calculation are presented in \figurename~\ref{fig:5}.
\begin{figure}
\begin{center}
\includegraphics[width=0.95\columnwidth,clip]{Fig5.eps}
\end{center}
\caption{Relaxation time of a single-nucleon excitation, $\tau_1$, as a function of the excitation energy $E_\text{ex}$. 
Calculations are shown for mass numbers $A = 50$, $150$, and $250$.}
\label{fig:5}
\end{figure}
As seen from the figure, unlike the behavior of the overall relaxation time $\tau$, the single-nucleon relaxation time $\tau_1$ increases with increasing excitation energy $E_\text{ex}$. This result is quite natural, since the farther the nucleon is from the Fermi surface, the longer it takes to reach equilibrium. Moreover, the dependence on 
the mass number $A$ is opposite to that of the overall relaxation time $\tau$: as $A$ increases, the relaxation time $\tau_1$ decreases. This can be easily explained by 
noting that the width of the single-particle peak is inversely proportional to the mass number [see Eq.~(\ref{p-width})]. That is, as $A$ increases, the peak width decreases, 
and consequently, the relaxation time shortens. As observed, $\tau_1$ is smaller than the overall relaxation time: $\tau_1 < \tau$. Hence, one can conclude that 
the relaxation time of an individual single-nucleon excitation is shorter than that of the nucleus as a whole. It is evident that the relaxation of the entire system 
occurs due to the faster single-particle processes, and therefore the overall relaxation time must exceed $\tau_1$.

Thus, the only factor that can influence the different types of relaxation times $\tau_n$ is the variation in the values of the kinetic coefficients, which determine 
the timescale of the evolution of the distribution function. To investigate the effect of this factor, one introduces a scaling divisor $y$, such that
$\widetilde{D}_{p,0} = D_{p,0}/y$ and $\widetilde{K}_{p,0} = K_{p,0}/y$. At the same time, their ratio, which determines the system temperature, is kept constant: 
$D_{p,0}/K_{p,0} = \widetilde{D}_{p,0}/\widetilde{K}_{p,0} = -T_\text{eq}$.

Figure~\ref{fig:6} shows the dependence of the overall relaxation time $\tau$ and the single-particle relaxation time $\tau_1$ on the introduced parameter $y$.
\begin{figure}
\begin{center}
\includegraphics[width=0.95\columnwidth,clip]{Fig6.eps}
\end{center}
\caption{Relaxation times of the system as a whole, $\tau$, and of the single-particle excitation, $\tau_1$, as functions of the parameter $y$.}
\label{fig:6}
\end{figure}
The calculation was performed for mass number $A = 150$ and excitation energy $E_\text{ex} = 10$ MeV. As seen, with increasing the parameter $y$, both $\tau$ and $\tau_1$ 
also increase. This indicates that as the diffusion and drift coefficients decrease, the relaxation rate slows down. Already with a 20-fold reduction of the kinetic 
coefficients, the value of $\tau$ reaches approximately $10^{-22}$ s. The relaxation time of the single-particle excitation, $\tau_1$, increases less steeply, but with 
a 200-fold reduction of the diffusion and drift coefficients, it reaches the order of $5 \times 10^{-23}$ s. Thus, achieving larger relaxation times is possible 
only by significantly reducing the values of the diffusion and drift coefficients. However, such values contradict the estimates obtained previously \cite{KoLu.IJMP.2015}. 
Therefore, in this work, one notes the presence of a discrepancy in determining the values of the diffusion and drift kinetic coefficients. 
Undoubtedly, this discrepancy may be influenced by the method of calculating the relaxation times; however, in my opinion, such large values of the parameter $y$ 
indicate that the method itself is not the main cause of the disagreement. Clearly, the explanation should be sought in other aspects of the model.

\section{Conclusions}

In summary, it should be noted that the main goal of this work was to investigate the time scales of single-particle excitation relaxation in a Fermi system, 
considered within the diffusion approximation of kinetic theory, in comparison with the typical two-particle relaxation time in the nucleus. 
For this purpose, the relaxation processes 
of a single-nucleon excitation in an atomic nucleus were analyzed within the diffusion approximation. The nucleus was modeled as a spherically symmetric Fermi system 
in both momentum and coordinate spaces. To describe the matter inside such a system, the approximation of infinite nuclear matter was employed, neglecting contributions 
from the system’s edge. The model was described by a diffusion equation with constant kinetic diffusion and drift coefficients, whose values were chosen based on previous 
studies such that their ratio, determining the system’s equilibrium temperature, remained physically reasonable.

The main idea of this work was to isolate the single-particle excitation from the steplike initial state of the core, against the background of which its relaxation occurs. 
In this context, three characteristic time scales were considered: the relaxation time of the entire system, the relaxation time of the step-like core distribution, 
and the relaxation time of the isolated single-particle excitation.

The calculations showed that, in all cases, the characteristic relaxation time is of the order of $\tau \approx 10^{-24}$~s, which is too short and does not agree with 
the typical two-particle relaxation time in the nucleus, $\tau_\text{coll} \approx 10^{-23} \div 10^{-22}$~s (Refs.~\cite{We.PRB.1974,Be.ZPA.1978,Ko.NPA.1982,ShKo.RPP.2004}). 
Isolating the single-particle excitation alone demonstrated that its relaxation occurs on the same time scale as that of the system as a whole. Furthermore, 
the characteristic time was found to be even shorter than in the case of the step-like distribution, which is natural, since it is precisely single-particle collisions 
that drive the relaxation of the system’s core.

The obtained dependencies of the relaxation time on the excitation energy reveal several characteristic patterns. In particular, the relaxation time of the entire system, 
$\tau \equiv \tau_\text{tot}$, decreases with increasing excitation energy $E_\text{ex}$, whereas the single-particle excitation exhibits the opposite behavior, 
with $\tau_1$ increasing. 
In both cases, the characteristic relaxation time grows with increasing mass number $A$; for the system as a whole, it gradually approaches the value typical for infinite 
nuclear matter, $\tau_{0} \approx 8.9 \times 10^{-24}$~s.

The relaxation time $\tau_{0}$ obtained in this work is smaller than the equilibrium time $\tau_\text{equ}$ reported in Ref.~\cite{Wo.PRL.1982}. This difference reflects 
the distinct definitions of the relaxation time: $\tau_{0}$ characterizes the system’s temporal evolution at finite times, whereas $\tau_\text{equ}$ corresponds to the 
asymptotic behavior of the exact analytical solution of the diffusion equation in the limit $t \to \infty$.

It was shown that the timescale of the relaxation processes is determined solely by the values of the diffusion and drift coefficients. Varying these coefficients allows 
one to examine the corresponding changes in the relaxation times. To this end, a scaling divisor $y$ was introduced, by which both coefficients were varied linearly while 
preserving their ratio. It was found that, as $y$ increases, the relaxation times $\tau$ and $\tau_1$ also increase. Hence, achieving longer relaxation times is possible 
only through a substantial reduction of the kinetic coefficients, although such values are inconsistent with previous estimates.

Thus, in this work, a discrepancy in the determination of the diffusion and drift coefficients has been noted. This inconsistency calls for further detailed analysis, 
in particular, a refinement of the assumptions underlying the diffusion approximation and an examination of the potential influence of quantum effects on the relaxation 
process.

\bigskip

\acknowledgments
The author thanks the Armed Forces of Ukraine for ensuring safety during this research. 
Computations were performed using the supercomputing cluster of the Bogolyubov Institute for Theoretical Physics, NAS of Ukraine. 

\section*{Data Availability Statement}
No additional data are available. All relevant data are contained within the article.

\bibliography{references}

\end{document}